\documentclass[preprint,authoryear,times,3p,12pt]{elsarticle}
\usepackage[T1]{fontenc}
\usepackage{geometry}
\usepackage{graphicx}
\usepackage{amssymb}
\usepackage{amsmath}
\usepackage{amsfonts}
\usepackage{multirow}
\usepackage{tabularray}
\usepackage{natbib}
\setcitestyle{authoryear,round}
\usepackage{url}
\usepackage{subcaption}

\usepackage{booktabs}
\usepackage{tikz}
\usetikzlibrary{positioning}
\usetikzlibrary{patterns}
\usetikzlibrary{arrows.meta}

\usepackage{tabularray}
\usepackage{mathtools}

\usepackage{epstopdf}
\usepackage{float}

\usepackage{etoolbox}

\newlength{\figwidthlarge} 			
\newlength{\figwidth} 			
\newlength{\figwidthsmall} 			
\usepackage{colortbl}

\journal{journal}

\begin{document}

\begin{frontmatter}

\title{From electricity prices to profits: multidimensional probabilistic forecasting for BESS trading}

\author[1]{Tomasz Weron\corref{cor1}} 
\ead{tomasz.weron@pwr.edu.pl}
\author[2]{Katarzyna Maciejowska}
\ead{katarzyna.maciejowska@pwr.edu.pl}

\affiliation[1]{
    organization={Department of Applied Mathematics, Wroc{\l}aw University of Science and Technology},
    country={Poland}
}

\affiliation[2]{
    organization={Department of Operations Research and Business Intelligence, Wroc{\l}aw University of Science and Technology},
    country={Poland}
}

\cortext[cor1]{Corresponding author: Tomasz Weron}

\date{June 19, 2026}

\begin{abstract}

This article examines various methods of constructing multidimensional probabilistic forecasts of electricity prices. Building on the Multiple Split (MS) method, it incorporates forecast averaging across estimation windows of different lengths and compares its performance with that of other, well-established methods. The research demonstrates that the ensemble representation of the price distribution is particularly useful in battery energy storage system (BESS) management. It enables the direct construction of probabilistic forecasts of daily profits. These forecasts can be used to determine optimal charging and discharging hours, as well as to support risk management decisions. The methods are evaluated using data from the German and Spanish day-ahead electricity markets from 2021-2024. The results indicate that the MS method with averaging (MS-ave) generally outperforms the other considered approaches in terms of Prediction Interval Coverage Probability (PICP), the Continuous Ranked Probability Score (CRPS) and the Energy Score (ES). Moreover, it is superior in supporting BESS trading strategies, particularly in case of non-zero operational costs.

\end{abstract}


\end{frontmatter}

\section{Introduction}
\label{sec:Introduction}

Over the last 20 years, electricity markets have experienced a continuous expansion of renewable energy sources (RES). In 2025, the capacity of RES increased by 16\%, with solar PV and wind accounting for over 75\% and 20\% of new renewable capacity additions worldwide, respectively~\citep{iea:26}. At the same time, conventional generation shifts from coal-fired plants to natural gas units. While these offer greater operational flexibility, they are also associated with higher and more volatile marginal production costs. Together with the continuous need to balance electricity supply and demand, these structural changes have increased electricity price volatility and exposed market participants to various forms of trading risk. 

In this context, grid-scale energy storage plays an important role in mitigating the market risk, as they enable the temporal shift of electricity consumption and generation. Nowadays, the most popular and mature technology is pumped-storage hydropower (PSH). However, its further expansion is constrained by site-specific geographical requirements. Therefore, battery energy storage systems (BESS) are emerging as a key alternative to PSH, due to their modularity and applicability for a wide range of locations. Although their installed capacity remains below that of PSH, BESS are expected to account for the majority of future storage expansion worldwide. The deployment of new BESS capacity increased from nearly 52~GW in 2022 to 108~GW by the end of 2025~\citep{eia:23, iea:26}, reflecting a rapid growth in recent years.

Reliable forecasts of electricity prices are essential for informed and rational participation in day-ahead electricity markets, whether by BESS operators or other market participants. However, increasing price volatility and generation uncertainty mean that point forecasts alone are insufficient for many operational and trading decisions. Effective risk management requires information about the range of possible future market outcomes and their associated probabilities. As a result, probabilistic forecasting has become an increasingly important tool for electricity market participants~\citep{now:wer:18} because, unlike point predictions, it approximates the entire distribution of the target variable. 

In the literature, three principal approaches dominate the construction of probabilistic forecasts. The first approach involves post-processing point forecasts to construct predictive distributions, often using quantile regression averaging~\citep{now:wer:15, uni:wer:21}. The second is based on the analysis of forecast errors, as in historical simulation or conformal prediction frameworks~\citep{kath_conformal_2021}. The third models the distribution of the target variable directly, for instance through GARCH-type specifications~\citep{jan:woj:22, bille:etal:2023} or distributional neural networks~\citep{mar:nar:wer:zie:23}. These approaches differ not only in the way uncertainty is estimated, but also in how it is represented. Depending on the framework, predictive uncertainty may be expressed through quantiles, ensembles of potential realizations, or parametric distribution functions. The choice of representation is particularly important when the objective is to forecast complex or non-linear processes, as some representations are more flexible and easier to apply than others.

In this research, we represent probabilistic forecasts in the form of ensembles rather than quantiles, which offers several advantages. First, unlike quantile-based methods, the ensemble approach can naturally represent forecasts of multidimensional random variables~\citep[see][as an example]{mac:nit:26}. While most forecasting methods, both point and probabilistic, focus on a single quantity, many real-world applications require the explicit modeling of dependencies among multiple variables~\citep{gne:etal:08, pin:13}. For instance, a wind farm operator may exploit the correlation between wind power generation and electricity price forecast errors to improve trading performance. Similarly, joint forecasting of electricity prices can support the development of trading strategies that optimize bids across multiple market segments, as demonstrated by~\cite{kumbartzky_optimal_2017}, \cite{mac:22}, \cite{jan:woj:22}. The ensemble representation also enables the straightforward construction of probabilistic forecasts for functions of multiple variables, such as the spread between intraday and day-ahead prices, or the revenue of a renewable energy producer participating in both short-term markets~\citep{mac:nit:26}. Such forecasts can be obtained by applying the function directly to each member of the ensemble, thereby generating an approximation of the corresponding probability distribution.

In this study, we apply the Multiple Split (MS) approach, proposed by~\cite{mac:nit:26}, to construct ensemble forecasts of a vector of twenty-four hourly electricity prices. To improve forecast accuracy, we extend the MS framework by incorporating forecast averaging. Specifically, we combine predictions generated by a common model specification, estimated using data from estimation windows of varying lengths. Forecast averaging across different window sizes has been proven successful in both point~\citep{mar:ser:wer:18,uni:mac:22} and probabilistic forecasting~\citep{ser:uni:wer:19,mac:uni:ser:24}. However, implementing such an extension within the MS framework is not straightforward, as it requires the coordination of the random splitting mechanism across different windows. To address this challenge, we propose a novel algorithm that ensures the resulting forecasts remain comparable and that their average has a meaningful interpretation.

Finally, we demonstrate how the MS method can be applied to probabilistic forecasting of BESS profits, which depend on the entire daily price curve rather than on individual hourly prices. Information about the predicted level of profits and the associated uncertainty is important for determining the optimal charging and discharging hours of the battery, and for deciding whether market conditions are favorable enough to justify trading. As noted by~\cite{hir:zie:26}, only forecasts that capture the joint distribution of prices across hours allow for a proper assessment of trading risk and consequently support optimal operational decisions. 

The proposed method is evaluated from three perspectives: the accuracy of electricity price forecasts, the quality of profit forecasts, and the economic value of predictions measured by BESS income from price arbitrage in the day-ahead market. While the existing literature has largely focused on the accuracy of electricity price forecasts, considerably less attention has been paid to the quality of forecasts of decision-relevant quantities derived from prices. In particular, to the best of our knowledge, the probabilistic forecasting of BESS profit distributions and the evaluation of their accuracy have not been systematically examined, which represents an important gap in the literature.

The empirical analysis is conducted using data from the German and Spanish electricity markets. Forecast accuracy is assessed using the prediction interval coverage probability (PICP) and the associated Kupiec test, the Continuous Ranked Probability Score (CRPS) and the Energy Score (ES). The proposed approach is subsequently compared with established benchmark methods, including the Quantile Regression Machine~\citep[QRM,][]{mar:uni:wer:20} and Historical Simulation~\citep[HS,][]{now:wer:18}.

The results demonstrate that the MS approach with forecast averaging (MS-ave) provides well-calibrated probabilistic forecasts of both electricity prices and BESS profits. The empirical coverage of the prediction intervals generated by the MS-ave method closely matches their nominal levels for both target variables. Moreover, in most cases, the MS and MS-ave outperform the benchmark approaches in terms of CRPS, both for the entire distribution and its tails, as well as with respect to the Energy Score.

While the existing literature has focused primarily on electricity prices, this study emphasizes the importance of probabilistic forecasting of BESS profits for operational decision-making. We demonstrate that more accurate forecasts of BESS income translate into economically significant gains. The superior accuracy of the MS-ave method leads to better trading decisions, resulting in higher profits and lower downside risk, as measured by the 5\% Value-at-Risk (VaR). These results highlight the importance of directly evaluating probabilistic forecasts of decision-relevant quantities, rather than focusing exclusively on the accuracy of forecasts of individual market variables.

The rest of this article is structured as follows. Section~\ref{sec:data} presents the main characteristics of the dataset explored in the study. Next, the forecasting methods are presented in Section~\ref{sec:methods}. Section~\ref{sec:simulation} describes the trading strategy of BESS. Finally, the results of the empirical study are presented in Section~\ref{sec:results}. Section~\ref{sec:conclusions} concludes the article.

\section{Data}
\label{sec:data}

\begin{figure}
    \centering
    \includegraphics[width=\textwidth]{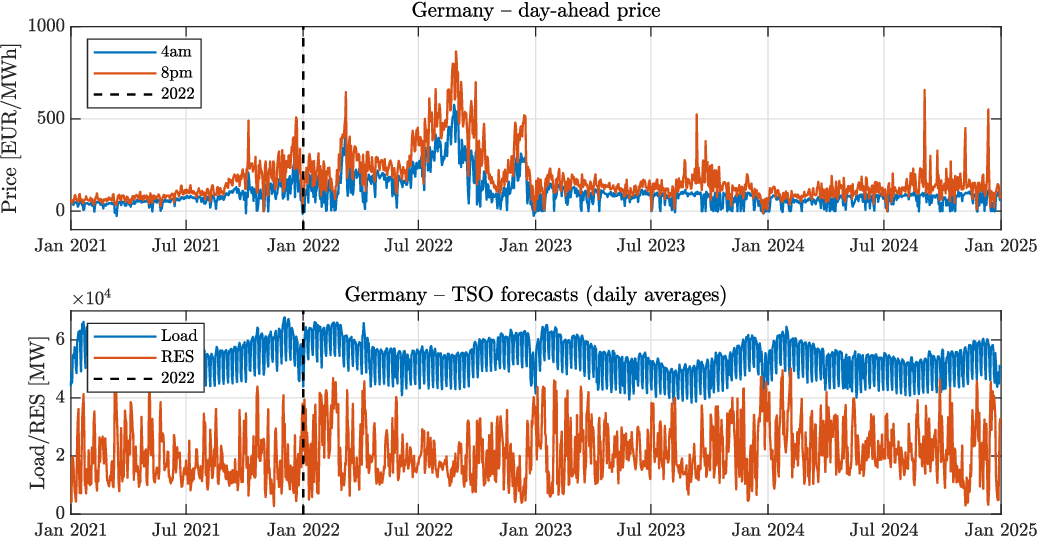}  
    \caption{German day-ahead market price for a selected off-peak (4am) and peak (8pm) hour (top), daily averages of forecasted load and renewable generation (RES, bottom). The black dashed line marks the beginning of the validation period.}
    \label{fig:ger}
\end{figure}

\begin{figure}
    \centering
    \includegraphics[width=\textwidth]{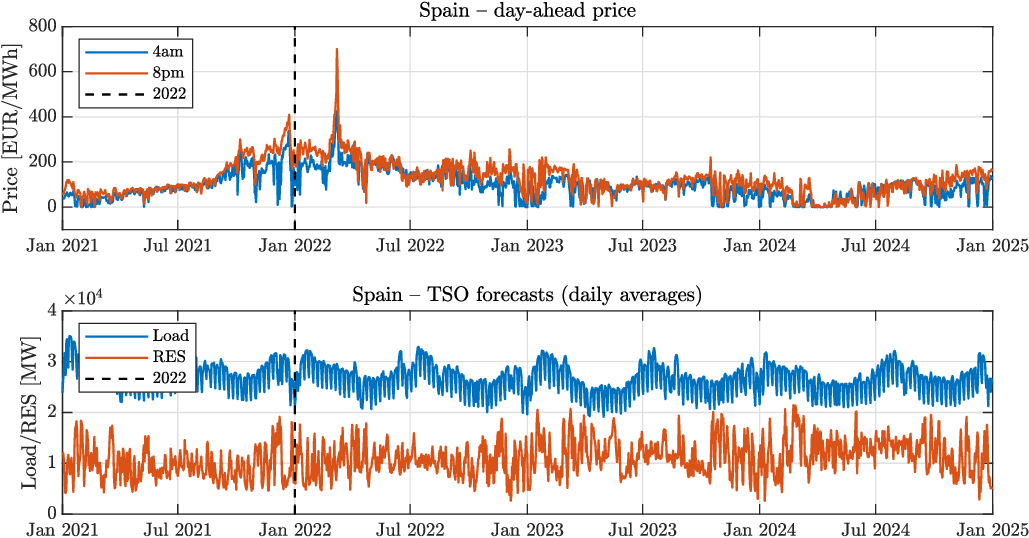}   
    \caption{Spanish day-ahead market price for a selected off-peak (4am) and peak (8pm) hour (top), daily averages of forecasted load and renewable generation (RES, bottom). The black dashed line marks the beginning of the validation period.}
    \label{fig:spa}
\end{figure}

\begin{figure}
    \centering
    \includegraphics[width=\textwidth]{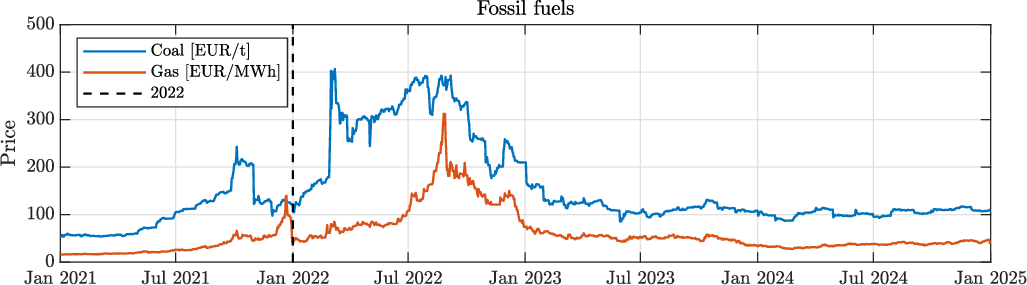}   
    \caption{Daily commodity futures prices, i.e. coal (API2) and gas (TTF). The black dashed line marks the beginning of the validation period.}
    \label{fig:fos}
\end{figure}

In the paper, we validate our models on two distinct electricity markets: German and Spanish. The data cover a period of four full years, from 1 January 2021 to 31 December 2024. The first year, 2021, is used purely for model calibration, while the subsequent years, 2022, 2023, and 2024, comprise a three-year long validation period. The dataset consists of day-ahead market prices ($DA$), accompanied by exogenous variables, i.e. forecasted load ($L$) and renewable generation (solar, wind offshore and onshore combined, $RES$), all of hourly resolution. The forecasts are the ones provided by the transmission system operators (TSO).

As shown in Figures~\ref{fig:ger}-\ref{fig:spa}, prices differ greatly across the two studied markets. Although both markets behaved similarly during the COVID-19 pandemic, the German market presents a much more articulated response to the Russo-Ukrainian war. In Spain, except for the initial stages of the war (the first half of 2022), prices seem unaffected by the ongoing conflict. To address the high volatility during the examined period, we include commodity futures prices for coal (API2, $C$) and natural gas (TTF, $G$) in daily resolution in the dataset (see Figure~\ref{fig:fos}).

The electricity market data used in this research is freely available at \url{transparency.entsoe.eu}, while the prices of coal and gas are sourced from \url{Investing.com}.

\section{Forecasting Methods}
\label{sec:methods}

In this study, we examine a day-ahead electricity market in which bids for all 24 hours are submitted around noon on the day preceding delivery. Consequently, hourly prices do not exhibit the typical structure of a time series. It is more appropriate to interpret them as distinct products rather than as sequential observations.

The forecasting framework considered in this paper is based on point forecasts derived from univariate models with a common specification for the entire day. To capture intraday heterogeneity, model parameters are estimated separately for each hour. Finally, to reflect the operational timing of trading in the day-ahead market, we assume that all computations are performed at 11:00~a.m. the day before delivery and restrict the information set accordingly to data available up to this time.

Probabilistic forecasts are derived from point predictions generated by autoregressive models with exogenous variables (ARX). These models are selected due to their computational efficiency, interpretability, and strong empirical performance. The proposed framework is flexible enough to accommodate alternative point forecasting methods, including classical time series models and modern machine learning approaches.

In this research, we represent probabilistic forecasts in the form of ensembles rather than quantiles, which offers several advantages. First, unlike quantile-based methods, the ensemble approach can be naturally extended to the construction of multidimensional forecasts of 24 hourly prices. In the considered frameworks, ensemble members are obtained as the sum of point predictions and forecast errors. When the forecast errors are constructed in a way that preserves the within-day correlation structure of the residuals, there is no need for further modeling of intraday dependencies of prices. 

The ensemble representation of the distribution also enables the straightforward computation of probabilistic forecasts for functions of electricity prices, such as the daily profits of a BESS. In this case, the function is applied directly to each member of the ensemble, yielding a direct representation of the corresponding distribution.
Finally, in the context of the Multiple Split forecasting method, the ensemble representation eliminates the need to average quantiles or prediction intervals when combining results obtained from different splits \citep{lei:etal:18, mac:nit:26}. As a result, the estimation procedure becomes both simpler and more accurate (see \cite{mac:nit:26} for a detailed discussion).

\subsection{Autoregressive Model}
\label{sec:model_arx}

In this research, we use an ARX model, which is a popular approach in the electricity price forecasting (EPF) literature~\citep[see][]{bille:etal:2023, lag:mar:des:wer:21}. The structure of the model is based on expert knowledge and literature, and is predefined \textit{a prior}.

Let us denote the electricity price on day $t$ and hour $h$ by $DA_{t,h}$. We adopt the following model specification:
\begin{align}
    DA_{t,h} =& \sum_{i=1}^7 D_i\theta_{i,h}
    + \sum_{i=1}^4 DA_{t-p_i} \theta_{7+i,h}
    + DA_{t-1,min} \theta_{12,h} + DA_{t-1,max} \theta_{13,h} \nonumber \\
    &+ L_{t,h} \theta_{14,h} + RES_{t,h} \theta_{15,h}
    + C_{t-2,h} \theta_{16,h} + G_{t-2,h} \theta_{17,h}
    + \varepsilon_{t,h}.
\label{eq:arx}
\end{align}
The model comprises four components: (i) seven week-days dummies; (ii) an autoregressive term with lags $p = \{1,2,3,7\}$, capturing short-term dynamics and weekly seasonality; (iii) minimum and maximum prices from previous day, $t-1$; (iv) exogenous variables -- forecasted load, $L$, and Renewable generation, $RES$, as well as coal $C$ and gas $G$ prices from day $t-2$. Calibration period is 364 days and the rolling window scheme is applied, i.e. the parameters, $\theta$, are estimated for each day $t$ separately. 

The ARX model specified above forms the basis for all subsequent methods, with the exception of the Na\"{\i}ve and Oracle strategies (discussed in Section \ref{ssec:strategies}).

\subsection{In-Sample Errors}
\label{ssec:model_is}

Let us assume that data is observed for periods $t = 1, \dots, T$, and that the objective is to forecast electricity prices for period $T+1$. The In-Sample (IS) method uses the entire sample to estimate the parameters of the model (\ref{eq:arx}). The parameters are then employed to compute both the in-sample forecasts, $\hat{DA}_{t,h}$, and the out-of-sample forecasts, $\hat{DA}_{T+1,h}$. The predictions are collected in the form of 24-dimensional vectors containing hourly quantities: $\hat{DA}_t = [\hat{DA}_{t,1}, \dots, \hat{DA}_{t,24}]'$. Next, the in-sample forecast errors are calculated as $e_t=DA_t-\hat{DA}_t$, where $DA_t=[DA_{t,1}, \dots, DA_{t,24}]'$. Finally,  the ensemble of predictions  is constructed as 
\begin{equation}
    \Psi^{(IS)}=\{Y\in \mathbb{R}^{24}: Y=\hat{DA}_{T+1} + e_t, t=1,\dots, T\}.
\end{equation}
Prediction intervals for individual hours are estimated using empirical quantiles of the forecasts in $\Psi^{(IS)}$.

It should be noted here  that the forecast errors, $e_t$, preserve the within-day correlation structure of the residuals; therefore, the resulting probabilistic forecast is inherently multidimensional. Consequently, it does not require imposing additional assumptions on the interdependencies between hours. When the objective is to predict a function of hourly prices -- for example, the daily profit of a battery energy storage system -- the corresponding probabilistic forecast can be directly obtained by applying the function to the individual elements of the ensemble.

The major disadvantage of IS approach is underestimation of the prediction risk. Since the same data is used to calibrate the model parameters and to calculate the errors, the resulting ensemble $\Psi^{(IS)}$ does not fully reflect the out-of-sample distribution of prices.

\subsection{Historical Simulation}
\label{ssec:model_hs}

Historical Simulation (HS) approach explores the out-of-sample forecast errors, which resemble better the distribution of future prices than in-sample errors discussed in the previous section. In this method, the data is split into two parts: a moving estimation window, used to calculate model parameters, and a calibration window, i.e. 182 last observations employed for construction of the probabilistic forecast~\citep{now:wer:18}. The algorithm consists of the following steps:
\begin{enumerate}
    \item The estimation window of initial $T_{estim}$ observations ($T_{estim}\leq T-182$) is used to estimate model parameters and to calculate the forecast $\hat{DA}_{T-181}$. Next, the estimation window is moved by one observation and the procedure is repeated. As the results, predictions for $t=T-181, \dots, T+1$ are computed.
    \item The forecast errors are estimated as $e_t = DA_{t}-\hat{DA}_t$ for $t=T-181, \dots, T$
    \item The ensemble of forecasts is constructed as 
    \begin{equation}
        \Psi^{(HS)}=\{Y\in \mathbb{R}^{24}: Y=\hat{DA}_{T+1} + e_t, t=T-181,\dots, T\}.
    \end{equation}
\end{enumerate} 

Notice that, unlike in the IS approach, the HS method uses only $182$ observations to construct the multidimensional distribution. Moreover, the model parameters are estimated using a relatively shorter estimation window of length $T-182$ at most. As a result, the forecasts are subject to additional estimation uncertainty.

\subsubsection{Forecast averaging in Historical Simulation}

In this article, similar to~\cite{hub:mar:wer:19}, we consider forecasts based on $M$ estimation windows of different lengths $T_{estim}\in\{T_1, \dots, T_M\}$, where $T_1< \dots <T_M=T-182$. In such a case, we keep the calibration window unchanged and use $T_m$ observations to fit the model parameters, to calculate forecast error, $e_t^{(m)}$, and the point prediction, $\hat{DA}_{T+1}^{(m)}$. The final ensemble takes the following form
\begin{equation}
    \Psi^{(HS-ave)}=\{Y\in \mathbb{R}^{24}: Y = \frac{1}{M}\sum_{m=1}^M\hat{DA}^{(m)}_{T+1}+\frac{1}{M}\sum_{m=1}^M e^{(m)}_t, t=T-181,\dots, T\}.
    \label{eq:HSave}
\end{equation}
%

\subsection{Quantile Regression Machine}
\label{ssec:model_qrm}

The Quantile Regression Machine (QRM) is a forecasting approach that links the distribution of future prices to their point predictions. It was introduced by \cite{ser:uni:wer:19} and, similar to Quantile Regression Averaging (QRA) approach of \cite{liu:now:hon:wer:17}, uses quantile regression (QR) to model and predict  quantiles of the target variables. It uses as an input the best available point forecast of DA prices, which is typically constructed as the average of predictions obtained from a single model estimated with samples of different lengths \citep{ser:uni:wer:19}.
 
{Let $Q_{\tau}(DA_{t,h})$ denote the $\tau$-quantile of the electricity prices $DA_{t,h}$. The QRM assumes that the quantile can be expressed as a linear function of the point forecast $\hat{DA}_{t,h}$
\begin{equation}
	Q_{\tau}(DA_{t,h}) 
	= \alpha_{\tau,h} + \hat{DA}_{t,h}\beta_{\tau,h},
	\label{eq:QRM}
\end{equation}
where $\alpha_{\tau,h}$ and $\beta_{\tau,h}$ are quantile- and hour-specific parameters. The parameters of Eq.~(\ref{eq:QRM}) are estimated by minimizing the sum of \emph{pinball scores} \citep[see][]{koe:hal:01}, which are defined as 
\begin{equation}\label{eq:pinball}
\mathrm{Pin}_{t,h}(\tau) =
\begin{cases}
(1-\tau)\bigl(Q_{\tau}(DA_{t,h}) - DA_{t,h}\bigr),  
& \text{if } DA_{t,h} < Q_{\tau}(DA_{t,h}), \\[0.2cm]
\tau\bigl(DA_{t,h} - Q_{\tau}(DA_{t,h})\bigr), 
& \text{if } DA_{t,h} \geq Q_{\tau}(DA_{t,h}).
\end{cases}
\end{equation}}

Since the QRM uses the forecasts, $\hat{DA}_{t,h}$, as inputs to the regression, the sample must be divided into estimation and calibration windows, similarly to the HS approach. Analogously to the HS, we assume that the calibration window consists of the last 182 observations, whereas the length of the estimation window may vary, such that $T_{estim} \leq T - 182$. When the forecast averaging is considered then $\hat{DA}_{t,h}$ is computed as an average of predictions based on estimation windows of different sized.

The model is estimated separately for 99 percentiles: $\tau = 0.01, \ldots, 0.99$, enabling the QRM to approximate the entire conditional distribution of $DA_{t,h}$. Since the quantiles are estimated independently, the resulting forecasts may violate the monotonicity condition: $Q_{\tau_1}(DA_{t,h}) \leq Q_{\tau_2}(DA_{t,h})$ for $\tau_1 < \tau_2$. To address this issue, the predicted quantiles are post-processed by sorting them to ensure monotonicity.

The QRM is designed to predict marginal distributions of hourly prices, and hence it does not provide any information about dependencies between hours. Therefore, in order to construct a multidimensional forecast, we assume that prices at different  hours are independent. In such a case, the multidimensional ensemble used for calculation of the profit distribution is constructed as the 
\begin{equation}
    \Psi^{(QRM)}=\{Y\in \mathbb{R}^{24}: Y_h= \hat{\alpha}_{\tau_h,h} + \hat{DA}_{T+1,h} \hat{\beta}_{\tau_h,h},\tau_h=0.01, \dots,0.99, h=1,\dots,24\},
\end{equation}
where $Y_h$ is the $h$-th element of vector $Y$ and the values of $\tau_h$'s are selected independently across hours. This implies that $\Psi^{(QRM)}$ includes all possible combinations of quantiles. If there are multiple estimation windows, $\Psi^{(QRM-ave)}$ is constructed analogously to $\Psi^{(HS-ave)}$, as in Eq.~(\ref{eq:HSave}). In empirical application, we consider only the pairwise distribution of prices, which reduces the number of elements in considered ensembles to $99^2$.

\subsection{Multiple Split}
\label{ssec:model_ms}

In this research, we employ the multiple split (MS) forecasting method introduced by \cite{mac:nit:26}. Similar to the split conformal prediction (SCP) method, also known as inductive conformal inference \citep{lei:etal:18, bar:etal:21, kath_conformal_2021}, the MS approach randomly divides the data into two disjoint subsets: estimation and calibration. The first (estimation) window is used to estimate the model parameters, which are subsequently employed to generate out-of-sample forecasts and to compute forecast errors for the observations in the calibration window. In the MS, the random splitting procedure is repeated $N$ times in order to improve forecast accuracy and reduce the variability of the results. The final ensemble is then constructed by directly aggregating the outputs obtained from the individual splits.

The detailed description of the MS algorithm is provided in \cite{mac:nit:26}. In this study, we employ a multidimensional version of the algorithm, in which all hourly prices are collected in a $(1 \times 24)$ vector ${DA}_{t}$. Suppose that the data-generating process can be represented by the following linear model:
\begin{equation}\label{eq:MS}
{DA}_{t} = Z_{t}A + e_{t},
\end{equation}
where $Z_{t}$ denotes a vector of exogenous variables, $A$ is a matrix of coefficients (with the $h$-th column corresponding to the $h$-th hour of the day), and $e_{t} \in \mathbb{R}^{24}$ is a vector of residuals. The parameters in $A$ may be estimated either jointly or equation-by-equation, depending on the model specification and the researcher’s preference. Moreover, the set of exogenous variables may differ across hours, as in (\ref{eq:arx}), which is equivalent to imposing zero restrictions on the matrix $A$.

\subsubsection{Forecast averaging in MS}

In this article, the MS method is extended to incorporate forecast averaging, which has been shown to improve predictive accuracy in both point and probabilistic forecasting settings~\citep{hub:mar:wer:19,ser:uni:wer:19}. Similar to \cite{hub:mar:wer:19}, we consider forecasts obtained with parameters calibrated to samples of different sizes. Suppose we consider $M$ estimation samples of different lengths, denoted by $S_1, \dots, S_M$, such that $T_1 < \dots < T_M \leq T$ and $S_m=\{t: T-T_m+1\leq t\leq T\}$.

{In order to construct ensembles that can be meaningfully averaged, the procedure must satisfy several conditions. First, the ensembles based on different samples, denoted by $\Psi_m$, must have equal sizes so that no specification is disproportionately represented. Second, they must be constructed using the same calibration windows to ensure that the resulting predictions are comparable and that their average has a clear interpretation. To achieve this, we propose the following algorithm. For each split:
\begin{enumerate}
    \item The shortest sample, $S_1$, is randomly split into two subsets $S_{estim}^{(1)}$ and $S_{calib}^{(1)}$ such that $S_1=S_{estim}^{(1)}\cup S_{calib}^{(1)}$ and $S_{estim}^{(1)}\cap S_{calib}^{(1)} = \emptyset$. The ensemble $\Psi_1$ is constructed as $\Psi_1=\{Y\in \mathbb{R}^{24}: Y = \hat{DA}^{(1)}_{T+1}+{e}^{(1)}_t, t\in S_{calib}^{(1)}\}$, where $\hat{DA}^{(1)}_{T+1}$ and ${e}^{(1)}_t$ are calculated using parameters estimated over $S_{estim}^{(1)}$.
    \item The longer samples are divided accordingly in such a way that $S_{calib}^{(m)} = S_{calib}^{(1)}$ and $S_{estim}^{(m)}=S_m/ S_{calib}^{(1)}$. Hence, the ensemble $\Psi_m$ becomes $\Psi_m=\{Y\in \mathbb{R}^{24}: Y = \hat{DA}^{(m)}_{T+1}+{e}^{(m)}_t, t\in S_{calib}^{(1)}\}$, where $\hat{DA}^{(m)}_{T+1}$ and ${e}^{(m)}_t$ are calculated using parameters estimated over $S_{estim}^{(m)}$.
    \item The final ensemble of predictions is constructed as
    \begin{equation}
        \Psi^{(MS-ave)} =\frac{1}{M}\sum_{m=1}^M \Psi_m= \{Y\in \mathbb{R}^{24}: Y = \frac{1}{M}\sum_{m=1}^M\hat{DA}^{(m)}_{T+1}+\frac{1}{M}\sum_{m=1}^M {e}^{(m)}_t, t\in S_{calib}^{(1)}\}.
    \end{equation}
\end{enumerate}}

\begin{figure}[ht]
    \centering
    \includegraphics[width=\textwidth]{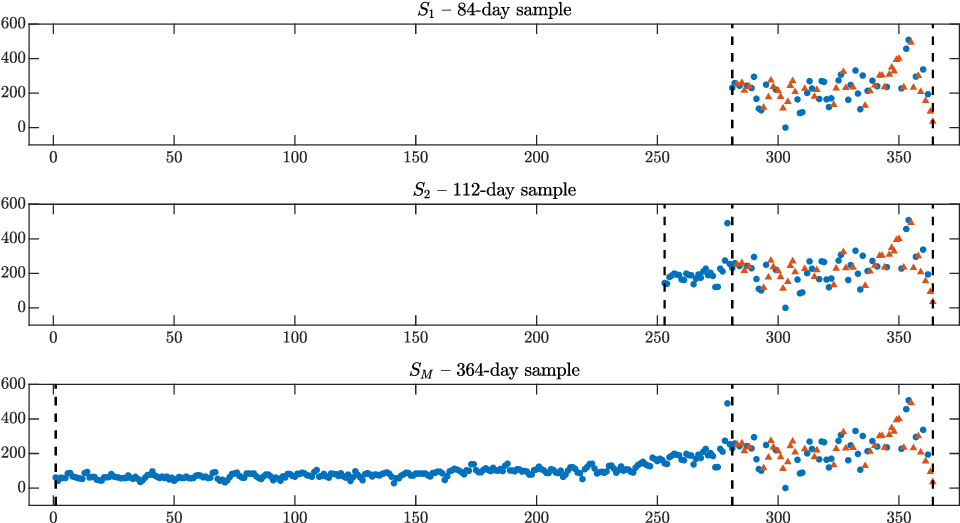}
    \caption{The data-splitting procedure in the Multiple Split (MS) method for estimation windows of different lengths.}
    \label{fig:cal_win}
\end{figure}

The data-splitting procedure is illustrated in Fig.~\ref{fig:cal_win}, which presents an example of how the samples are divided into estimation and calibration windows. The top panel shows the shortest sample of length $T_1 = 84$, for which all observations are randomly assigned either to $S_{estim}^{(1)}$ (blue circles) or to $S_{calib}^{(1)}$ (red triangles). We assume that both subsets are of equal size; hence, $|S_{estim}^{(1)}| = |S_{calib}^{(1)}| = 42$. Next, as the sample size increases, the estimation window is extended by concatenating additional observations. Therefore, for sample $S_2$ with $T_2 = 112$, the size of the calibration window remains unchanged, i.e. $|S_{calib}^{(2)}| = 42$, whereas the size of the estimation window increases to $|S_{estim}^{(2)}| = 70$. Finally, for the largest considered sample, which contains one year of observations ($T_M = 364$), the length of the estimation window becomes $|S_{estim}^{(M)}| = 322$.

The entire algorithm is repeated $N$ times, and the resulting ensembles are concatenated across iterations. Since a single iteration produces $T_1/2 = 42$ predictions, the final ensemble consists of $N \cdot T_1/2$ forecasts. These forecasts are subsequently used either to construct prediction intervals for each hour separately -- in which case the marginal distributions are considered -- or to predict the distribution of battery profits, thereby exploiting the multidimensional nature of the MS approach.

\subsection{Specification of forecasting methods}

In this research, we consider four approaches for constructing probabilistic forecasts: IS, HS, QRM, and MS. Additionally, the last three methods are extended to allow for forecast averaging across different estimation windows. The resulting specifications are denoted by HS-ave, QRM-ave, and MS-ave. Notice that HS, QRM and MS are the special cases of HS-ave, QRM-ave  and MS-ave, which include predictions based on a single estimation window.

The considered methods differ in terms of data usage and the size of the resulting multidimensional ensembles. We assume that all approaches explore one year of observations to construct probabilistic forecasts. In the IS method, the entire sample is used both to estimate the model parameters and to construct the ensemble. In the remaining methods, the data are divided into estimation and calibration windows. In HS, HS-ave, QRM and QRM-ave, the last 182 observations are used to calibrate the forecast distribution. In MS without forecast averaging, the calibration window has the same size, i.e., 182 observations, but the splitting procedure is repeated multiple times. Consequently, the forecast errors used to construct probabilistic forecasts collectively represent the entire sample. Finally, in MS-ave, only  $T_1/2$, here equal to 42 observations, are used to construct the ensemble within a single split. A summary of the considered selections of calibration windows is presented in Table~\ref{tab:specification}.

The choice of the calibration window also affects the amount of data available for parameter estimation. For methods without forecast averaging, all remaining observations are used to estimate the model parameters. For setups with forecast combination, however, the sizes of the estimation windows differ across approaches (see Table \ref{tab:specification}). In the MS framework, we consider two short and two long subsamples with $|S_1|=84$, $|S_2|=112$, $|S_3|=357$, and $|S_4|=364$, resulting in estimation windows of sizes $|S_{estim}^{(1)}| = 42$, $|S_{estim}^{(2)}| = 70$, $|S_{estim}^{(3)}| = 315$, and $|S_{estim}^{(4)}| = 322$, respectively. However, in the HS and QRM, the estimation window cannot exceed 182 observations. Here, we consider windows of lengths $T_1=42$, $T_2=56$, $T_3=178$, and $T_4=182$, corresponding to half of the subsamples $S_1, \dots, S_4$. Notice that, for all approaches, the longest estimation window complements the calibration window to the full one-year sample.

Substantial differences between the methods can also be observed in terms of the ensemble sizes. In the simplest case of the IS, $|\Psi^{(IS)}|=364$, whereas for the HS and HS-ave the ensembles consist of 182 predictions. The size of $\Psi^{(QRM)}$ and $\Psi^{(QRM-ave)}$ is the largest and reaches $99^{24}$ due to the assumption of independence across hours. However, in the empirical application we consider only the bi-hourly distribution, which reduces the number of ensemble elements to $99^2$. Finally, in the MS approach, the size of $\Psi^{(MS)}$ depends on the number of splits and equals $182 \cdot N$, which is substantially larger than in the IS and HS approaches. For the MS-ave specification, the ensemble size depends on the length of the shortest window used for forecast averaging. When $T_1=84$, then $|\Psi^{(MS-ave)}|=42 \cdot N$, and it increases proportionally with the number of splits.

\begin{table}
\caption{Comparison of forecasting methods}
\label{tab:specification}
\centering
\begin{tblr}{
colspec={|l|c|c|c|},
cell{1}{2} = {c=3}{c},
cell{1}{1} = {r=2}{m}
}
\hline
Method & Size of & & \\
\hline
& Estimation window(s) & Calibration window & Ensemble\\
\hline\hline
IS & 364 & 364 & 364 \\
\hline
HS & 182 & 182 & 182\\
QRM & 182 & 182 & 99$^{24}$ \\
MS & 182 & 182 & $182 \cdot N$ \\
\hline
HS-ave & 42, 56, 178, 182 & 182 & 182\\
QRM-ave & 42, 56, 178, 182 & 182 & 99$^{24}$ \\
MS-ave & 42, 70, 315, 322 & 42 & $42 \cdot N$\\
\hline \hline
\end{tblr}

\vspace{0.2cm}
\small{Note: for the QRM and QRM-ave approaches, the size of the ensemble is calculated for a bi-hourly distribution; $N$ denotes the number of random splits in the MS method.}
\end{table}

\subsection{Forecast Evaluation}
\label{ssec:forecast_eval}

The forecast accuracy of the proposed methods is evaluated from two perspectives: their ability to predict the marginal distributions of hourly electricity prices and the accuracy of probabilistic forecasts of daily BESS profits. In the latter case, we consider two alternative specifications. First, it is assumed that the battery is charged and discharged at predefined hours, here hours 4 and 20, respectively. In this setting, the profit forecast is a function of only two electricity prices and is therefore based on their joint bi-hourly distribution. Second, we consider all feasible pairs of charging and discharging hours, denoted by $h^{\mathrm{ch}}$ and $h^{\mathrm{dis}}$, that satisfy the operational constraint $h^{\mathrm{ch}} < h^{\mathrm{dis}}$. For each pair, a probabilistic forecast of BESS profits is constructed and evaluated. The reported results correspond to the average values of the accuracy measures across all 276 admissible charging-discharging combinations.

We use multiple measures to evaluate the quality of forecast. The first two are based on predictions of distribution quantiles. First, we assess the accuracy of prediction intervals with nominal coverage level of 95\% using the \emph{prediction interval coverage probability} (PICP). It measures the frequency with which observations fall within the predicted intervals in the testing window. We calculate the proportion for each hour separately ($\text{PICP}_h$) and then take the average over the whole evaluation period (PICP).

To formally assess whether the empirical (PICP) and nominal coverage levels are consistent, we apply the Kupiec test \citep{kup:95}. The test is performed separately for each hour, and the average proportion of hours for which the null hypothesis cannot be rejected is reported.

Finally, to assess the quality of the estimated quantiles, we calculate Continuous Ranked Probability Score (CRPS) proposed by \cite{gne:bal:raf:07}, which is proportional to the average level of \emph{pinball score} (see Eq.~(\ref{eq:pinball})) across 99 quantiles. For a given day $t$ and hour $h$, it is defined as
\begin{equation}
    \mathrm{CRPS}_{t,h}^{(99)} \approx \frac{2}{99}\sum_{\tau=0.01}^{0.99} \mathrm{Pin}_{t,h}(\tau).
\end{equation}

Additionally, to  assess the ability to approximate the tails of the distribution, we use a tail-focused variant of the CRPS, i.e. 
\begin{equation}
    \mathrm{CRPS}_{t,h}^{(20)} 
    \approx \frac{2}{20}\left(
    \sum_{\tau=0.01}^{0.1} \mathrm{Pin}_{t,h}(\tau)
    + \sum_{\tau=0.90}^{0.99} \mathrm{Pin}_{t,h}(\tau)
    \right),
\end{equation}
which averages the \emph{pinball scores} over the extreme lower and upper quantiles only. The final measures, $\mathrm{CRPS}_{99}$ and $\mathrm{CRPS}_{20}$, are obtained by averaging  $\mathrm{CRPS}^{(99)}_{t,h}$ and $\mathrm{CRPS}^{(20)}_{t,h}$ over the entire evaluation window.

For multidimensional random variables, $DA_t\in \mathbb{R}^{24}$, the quality of probabilistic forecasts may be also evaluated using the Energy Score (ES; \cite{gne:bal:raf:07}).
For ensemble forecasts, the ES can be directly approximated by
\begin{equation}
   \mathrm{ES} = \frac{1}{N}\sum_{i=1}^N ||DA_{t}-\hat{DA}_{t}^{(i)}|| - \frac{1}{2}\frac{1}{N^2}\sum_{i=1}^N\sum_{j=1}^N ||\hat{DA}_{t}^{(i)}-\hat{DA}_{t}^{(j)}||, 
\end{equation}
where $\hat{DA}_t^{(i)}$ and $\hat{DA}_t^{(j)}$ are the \textit{i}-th and \textit{j}-th elements of the ensemble and the distance between vectors is measured with L1 norm: 
\begin{equation}
 ||\hat{DA}_{t}^{(i)}-\hat{DA}_{t}^{(j)}|| = \sum_{h=1}^{24} |\hat{DA}_{t,h}^{(i)}-\hat{DA}_{t,h}^{(j)}|. 
\end{equation}
%

\section{Decision support of BESS}
\label{sec:simulation}

In this study, we demonstrate how the joint prediction of electricity prices across consecutive hours can support the operational decision-making of an energy storage system. Specifically, we analyze the trading strategy of a battery energy storage system (BESS) with an energy capacity of $1\,\mathrm{MWh}$ and a C-rate of one, implying that the battery can be fully charged or discharged within a single hour. Although BESSs can participate in multiple electricity markets, we restrict our analysis to price arbitrage in the day-ahead market. The objective of the BESS is to purchase electricity during low-price hours and sell it during high-price hours later the same day.

Operational planning requires the BESS to determine, one day in advance, the optimal timing of charging and discharging. These decisions are based on forecasts of day-ahead electricity prices. Let $h^{\mathrm{ch}}$ and $h^{\mathrm{dis}}$ denote the selected hours for charging and discharging, respectively. The profit obtained on day $t$ is then given by
\begin{equation}\label{eq:profits}
\pi_t = (1-\eta) DA_{t,h^{\mathrm{dis}}} - (1+\eta) DA_{t,h^{\mathrm{ch}}} - \mathrm{Cost},
\end{equation}
where $\eta$ denotes the round-trip efficiency of charging and discharging and $\mathrm{Cost}$ captures variable operational costs that have a direct impact on the decision process.
Although BESS does not require fossil fuels to generate and sell electricity, it is treated as a consumer by the power grid and is therefore subject to distribution fees for all electricity purchased from the grid. In 2024, average grid fees in Germany amounted to approximately $4\,\mathrm{ct/kWh}$ for energy-intensive industries, $9\,\mathrm{ct/kWh}$ for other commercial consumers, and about $11\,\mathrm{ct/kWh}$ for households, which correspond to costs of $40\,\mathrm{EUR/MWh}$, $90\,\mathrm{EUR/MWh}$ and $110\,\mathrm{EUR/MWh}$, respectively.

The utility makes its trading decision in the day-ahead stage, when the realized electricity prices are not yet known. Consequently, the decision process can rely only on information derived from forecasted profits,
\begin{equation}\label{eq:profits:forecast}
\hat{\pi}_t
= (1-\eta) \hat{DA}_{t,h^{\mathrm{dis}}} - (1+\eta) \hat{DA}_{t,h^{\mathrm{ch}}} - \mathrm{Cost},
\end{equation}
where $\hat{DA}_{t,h}$ denotes the forecast of the day-ahead electricity price for hour $h$ on day $t$, computed before noon on the day preceding delivery.

\subsection{Trading strategies}\label{ssec:strategies}

\subsubsection{Na\"\i ve strategy}

In this research, we consider a Na\"{\i}ve trading strategy that assumes fixed charging and discharging hours. Specifically, the utility is assumed to charge the battery at 4:00~a.m. and discharge it at 8:00~p.m., which typically correspond to the lowest and highest electricity prices within a day, respectively.

\subsubsection{Point forecast strategies}

Instead of assuming fixed charging and discharging hours, BESS may base its decisions on point forecast of electricity prices. In such a case, for each pair of hours, a point prediction of profits is calculated according to Eq.~(\ref{eq:profits:forecast}). The optimal charging and discharging hours are selected to maximize the expected income. This strategy, without any stopping rule, is referred to in the remainder of this research as ARX. Since for some days, the price spread may not be sufficient to cover operation cost, we consider a variant in which BESS stops operating when $\hat{\pi}_t<0$ (referred to as ARX-s).

\subsubsection{Probabilistic forecast strategies}
\label{sssec:sim_prob}

Similar to point forecast strategies, strategies exploiting probabilistic forecasts of prices consists of two steps. First, for each pair of hours $h_1 < h_2$, we construct an ensemble of profits under the assumption that the BESS is charged at hour $h_1$ and discharged at hour $h_2$. Next, we select the pair characterized by the highest median of income. Finally, the BESS is allowed to refrain from trading when the risk associated with the operation exceeds a predefined threshold.

In this article, risk is measured by the probability of incurring financial losses. We assume that the BESS neither buys nor sells electricity when the probability $\mathrm{prob}(\hat{\pi}_t < 0)$ exceeds a threshold $q$. As the result, the level of daily profits depends on a choice of $q$ and can be calculated as
\begin{equation}
\pi_t(q) = \begin{cases}
\pi_t,  
& \text{if }  \mathrm{prob}(\hat{\pi}_t < 0) < q, \\[0.2cm]
0, 
& \text{if }\mathrm{prob}(\hat{\pi}_t < 0) \geq q.
\end{cases}
\end{equation}
The value of the threshold $q$ reflects the trader's risk tolerance. A risk-averse manager will typically stop trading even when the probability of losses is relatively low, e.g., for $q = 0.3$. In contrast, a risk-neutral strategy may rely on intermediate values close to $q = 0.5$. 

It should be noted that the stopping condition $\mathrm{prob}(\hat{\pi}_t < 0) \geq q$ is equivalent to the rule $Q_{q}\!\left(\hat{\pi}_{t}\right) < 0$ proposed in \cite{mac:nit:26}. Using the example of a renewable energy source (RES) generator, \cite{mac:nit:26} showed that refraining from trading when operational risk is high may not only reduce Value at Risk (VaR), but also increase income level.

\subsubsection{Oracle}

The performance of the proposed trading strategies is benchmarked against the Oracle strategy that assumes perfect foresight of future electricity prices. Information about  prices is exploited in two ways. First, it is used to select the optimal charging and discharging hours. Second, BESS decides whether to engage in trading based on realized profits, $\pi_t$. If future profits are negative, the BESS refrains from operation and waits for more favorable market conditions.

\subsection{Economic evaluation of strategies}

{The performance of the proposed trading strategies is evaluated in terms of profitability, trading risk, and trading frequency. The total profit is calculated as the sum of individual daily profits:
\begin{equation}
\pi(q) = \sum_{t=1}^T \pi_{t}(q).
\end{equation}}

To better understand the role of the stopping rule in profit generation, we compare the results obtained for $q \in {0.1, \dots, 0.9}$ with those of a strategy that assumes unconditional trading, which is equivalent to setting $q = 1$. Additionally, we calculate the trading frequency, which indicates how often the stopping condition is not violated.

Finally, the trading risk of the strategies is evaluated using the Value-at-risk, $\mathrm{VaR}_{5\%}$, defined as the 5\% quantile of the profit distribution $\pi_{t}(q)$. Notice that under unfavorable market conditions or high operating costs, there may be many days with zero profits due to the high estimated risk of losses. Therefore, $\mathrm{VaR}_{5\%}$ is computed only for days on which trading actually occurs.

\section{Results}
\label{sec:results}

The forecasting experiment is based on a rolling window scheme (as mentioned in Section~\ref{sec:model_arx}) in which one year of observations is used to construct probabilistic forecasts of 24 hourly prices. The results are evaluated over the three-year validation period (see Section~\ref{sec:data} for details).

\subsection{Forecast accuracy of electricity prices}
\label{ssec:res_price}

\begin{table}[ht]
\caption{Forecast accuracy of electricity prices -  German and Spanish markets.}
\label{tab:price}
\centering
\begin{tblr}{
colspec={|l|cc|cc|c|},
cell{2}{1} = {c=6}{c},
cell{10}{1} = {c=6}{c},
}
\hline
& PICP$_{95\%}$ & Kupiec test & CRPS$_{99}$ & CRPS$_{20}$ & ES \\
\hline \hline
Germany & & & & & \\
\hline \hline
IS	&	92.64\%	&	8.33\%	&	18.666	&	8.514	&	18.473	\\
\hline
HS	&	94.45\%	&	95.83\%	&	17.800	&	8.102	&	17.639	\\
QRM	&	93.43\%	&	45.83\%	&	17.217	&	7.694	&	17.063	\\
MS	&	\textbf{95.43\%	}&	91.67\%	&	18.313	&	7.986	&	18.142	\\
\hline
HS-ave	&	94.44\%	&	91.67\%	&	16.742	&	7.756	&	16.591	\\
QRM-ave	&	93.55\%	&	45.83\%	&	\textbf{16.231}	&	7.399	&	\textbf{16.085}	\\
MS-ave	&	94.29\%	&	91.67\%	&	16.421	&	\textbf{6.999}	&	16.268	\\

\hline \hline
Spain & & & & & \\
\hline \hline
IS	&	93.28\%	&	25.00\%	&	12.848	&	5.562	&	12.714	\\
\hline
HS	&	\textbf{94.92\%}	&	100.00\%	&	12.929	&	5.407	&	12.807	\\
QRM	&	93.52\%	&	37.50\%	&	12.837	&	5.447	&	12.719	\\
MS	&	95.51\%	&	91.67\%	&	12.705	&	5.358	&	12.585	\\
\hline
HS-ave	&	94.74\%	&	100.00\%	&	12.179	&	5.207	&	12.065	\\
QRM-ave	&	93.79\%	&	54.17\%	&	12.054	&	5.203	&	11.943	\\
MS-ave	&	94.18\%	&	75.00\%	&	\textbf{12.016}	&	\textbf{5.030}	&	\textbf{11.902}	\\
\hline
\end{tblr}
\end{table}

Let us first evaluate the accuracy of probabilistic forecasts of electricity prices. The results for both markets -- German and Spanish -- are presented in Table~\ref{tab:price}, which reports the PICP with a nominal coverage level of 95\%, together with the outcome of the corresponding Kupiec test, CRPS$_{99}$, CRPS$_{20}$ and ES for each modeling approach.

First, it can be observed that almost all methods produce prediction intervals that are slightly too narrow, although the deviations from the nominal levels are generally moderate. Among the considered approaches, IS, QRM, and QRM-ave exhibit the greatest discrepancies between empirical and nominal coverage levels, which is confirmed by the frequent rejection of the null hypothesis in the Kupiec test. In the case of the IS method, the Kupiec test confirms correct empirical coverage in only 8.3\%--25\% of cases. At the same time, the HS- and MS-based approaches provide better-calibrated prediction intervals, with PICPs close to the nominal levels, and the Kupiec test indicating correct coverage in most cases. For instance, for the MS method, the null hypothesis in the Kupiec test is not rejected in 91.7\% of cases.

Although correct coverage is an important indicator of forecast accuracy, it does not evaluate the entire distribution and does not account for forecast sharpness. Therefore, we additionally report CRPS measures, which summarize the accuracy of forecasts across a range of quantiles. The results confirm that the IS approach provides the least accurate predictions, with CRPS being either the highest or the second highest, depending on the market. Unlike in the case of PICP, the HS method is outperformed by both the QRM- and MS-based approaches. Finally, the MS-ave method achieves the lowest CRPS values overall, being outperformed only by the QRM-ave in terms of CRPS$_{99}$ in the German market. Values of ES are strongly correlated with CRPS and confirm the outcomes of the latter.

When models with and without forecast averaging are compared, the results indicate that aggregating information from different forecasts improves predictive accuracy, particularly in terms of CRPS$_{99}$ and CRPS$_{20}$. An improvement can be observed for each modeling approach, with reductions ranging from 5.4\% to 10.3\% in CRPS$_{99}$ and from 3.7\% to 12.4\% in CRPS$_{20}$. The largest increase in accuracy is observed for the MS-ave method in the German market, where CRPS$_{99}$ and CRPS$_{20}$ decrease from 18.313 and 7.986 to 16.421 and 6.999, respectively. Once more, values of ES exhibit a similar pattern as CRPS.

Finally, a comparison of the CRPS and ES levels in Germany and Spain confirms the different behaviour of electricity prices across these two markets. As illustrated in Figures~\ref{fig:ger}--\ref{fig:spa}, not only are electricity prices in Germany higher on average, they are also substantially more volatile, making forecasting a more challenging task.

\subsection{Forecast accuracy of profits}
\label{ssec:res_profit}

Next, we evaluate the accuracy of probabilistic forecasts of BESS profits. The results for the two analyzed markets are presented in Table~\ref{tab:profit}, which reports PICP and CRPS measures. As described in Section~\ref{sec:methods}, forecast quality is assessed from two perspectives. First, we consider a fixed charging-discharging schedule, with charging at hour 4 and discharging at hour 20 (left panel of Table~\ref{tab:profit}). Second, we evaluate forecasts across the full set of feasible charging-discharging pairs satisfying $h^{\mathrm{ch}} < h^{\mathrm{dis}}$ and report the average values of the accuracy measures over all admissible combinations (right panel of Table~\ref{tab:profit}).

The quality of profit forecasts depends on two factors: the accuracy of the marginal distributions of electricity prices and the ability to capture within-day dependencies across hours. This is well illustrated by the performance of the IS and QRM approaches. As noted earlier, the IS method tends to underestimate forecasting uncertainty for individual hourly prices (see Table~\ref{tab:price}). As a consequence, it produces prediction intervals for BESS profits that are too narrow, with empirical coverage rates ranging from 91.06\% to 93.98\%, below the nominal level. By contrast, the QRM-based methods provide relatively well-calibrated marginal forecasts, but fail to capture the dependence structure between prices observed at different hours. Consequently, to construct the ensemble, we assume independence across hours, which leads to an overestimation of forecasting uncertainty. As a result, the empirical coverage of the 95\% prediction intervals generated by the QRM and QRM-ave approaches ranges from 96.44\% to 98.42\%.

Finally, it can be observed that the MS-based approaches outperform the other methods, as they provide well-calibrated marginal distributions while preserving the natural correlation structure of forecast errors. In particular, the MS provides prediction intervals with empirical coverage rates closest to their nominal levels in the German market, while the MS-ave does so in the Spanish market. At the same time, the MS-ave achieves the lowest values of the CRPS measure, both when calculated across all 99 quantiles and when focused exclusively on the tails of the distribution. These findings are remarkably consistent across the two analyzed electricity markets and under both evaluation settings. 

\begin{table}[ht]
\caption{Forecast accuracy of profits -  German and Spanish markets.}
\label{tab:profit}
\centering
\begin{tblr}{
colspec={|l|c|cc|c|cc|},
cell{1}{2,5} = {c=3}{c},
cell{3}{1} = {c=7}{c},
cell{11}{1} = {c=7}{c},
}
\hline
& Hours: 4 - 20 & & & Hours: all & & \\
\hline
& PICP$_{95\%}$ &  CRPS$_{99}$ & CRPS$_{20}$ & PICP$_{95\%}$ &  CRPS$_{99}$ & CRPS$_{20}$ \\
\hline \hline
Germany & & & & & & \\
\hline \hline
IS	&	91.06\%	&	19.805	&	9.211	&	91.55\%	&	15.673	&	7.165	\\
\hline
HS	&	93.70\%	&	20.269	&	9.239	&	93.81\%	&	15.668	&	7.108	\\
QRM	&	96.62\%	&	19.963	&	9.088	&	97.56\%	&	16.456	&	8.177	\\
MS	&	\textbf{94.80\%}	&	19.533	&	8.766	&	\textbf{95.21\%}	&	15.351	&	6.690	\\
\hline
HS-ave	&	92.88\%	&	20.240	&	9.672	&	93.63\%	&	15.486	&	7.311	\\
QRM-ave	&	96.44\%	&	19.438	&	8.969	&	97.26\%	&	16.084	&	8.063	\\
MS-ave	&	94.71\%	&	\textbf{18.617}	&	\textbf{8.176}	&	94.70\%	&	\textbf{14.321}	&	\textbf{6.248}	\\
\hline \hline
Spain & & & & & & \\
\hline \hline
IS	&	93.98\%	&	13.694	&	5.905	&	92.71\%	&	10.921	&	4.664	\\
\hline
HS	&	94.07\%	&	13.523	&	5.797	&	94.26\%	&	11.133	&	4.652	\\
QRM	&	97.63\%	&	13.803	&	6.159	&	98.42\%	&	11.917	&	5.670	\\
MS	&	96.17\%	&	13.588	&	5.801	&	95.65\%	&	10.817	&	4.516	\\
\hline
HS-ave	&	94.07\%	&	13.534	&	5.705	&	94.30\%	&	11.216	&	4.732	\\
QRM-ave	&	97.63\%	&	13.541	&	5.916	&	98.10\%	&	11.736	&	5.502	\\
MS-ave	&	\textbf{95.53\%}	&	\textbf{12.973}	&	\textbf{5.459}	&	\textbf{94.95\%}	&	\textbf{10.588}	&	\textbf{4.396}	\\
\hline
\end{tblr}
\end{table}

\subsection{Economic Evaluation}
\label{ssec:res_trade}

Finally, we examine the financial gains from using probabilistic forecasts in BESS management.  The results for the German and Spanish markets are presented in Tables~\ref{tab:trade_ger} and~\ref{tab:trade_spa}, respectively. We report the optimal risk threshold, denoted by $q^*$, that maximizes profits (for strategies incorporating a stopping rule), the trading frequency, the total profit, and the associated VaR$_{5\%}$. Additionally, we consider two levels of variable costs: 0 EUR/MWh and 40 EUR/MWh, reflecting  the transaction costs of BESS (see Eq.~(\ref{eq:profits})).

In the top panels of Tables~\ref{tab:trade_ger} and~\ref{tab:trade_spa}, we show the results for the Oracle strategy, which assumes perfect knowledge of future prices and allows the BESS to refrain from operation whenever losses would occur. The reported values correspond to the aggregated profit over the three-year evaluation period for a battery with a total capacity of 1 MWh. The profit levels obtained by the remaining strategies are expressed relative to the Oracle benchmark, thereby illustrating the loss of income resulting from imperfect forecasts and suboptimal decision rules.

Let us first note that, in the absence of variable costs, the Oracle strategy engages in trading on more than 99\% of days, indicating that the within-day price spread is typically sufficient to compensate for energy losses incurred during the charging--discharging cycle. At the same time, substantial differences in profitability can be observed across countries, with total profits in Spain being 42.7\% lower than those in Germany. When variable costs of 40 EUR/MWh are taken into account, the trading frequency decreases to 80.4\% and 57.5\% in Germany and Spain, respectively. This indicates that, on many days, trading revenues are insufficient to cover operating costs. Additionally, profit levels decline by 38.4\% in Germany and 59.8\% in Spain, further widening the profitability gap between these two markets.

Next, we consider the results of strategies that either assume fixed charging and discharging hours (Na\"{\i}ve) or select these hours using point forecasts, with (ARX-s) or without (ARX) a stopping rule. The results indicate a relatively poor performance of the Na\"{\i}ve approach, which achieves only 49.9\% (Germany) and 11.0\% (Spain) of the Oracle profits in the zero-cost setup. At the same time, the point forecast strategies yield profit levels close to or exceeding 90\% of the Oracle benchmark. Moreover, it can be observed that the stopping rule gains importance as operating costs rise. Under higher costs, the ARX-s strategy becomes more profitable than ARX, with the difference in profits being more pronounced in the Spain than in Germany. Finally, refraining from operation not only increases profitability but also reduces trading risk, as measured by VaR$_{5\%}$ (where a higher VaR$_{5\%}$ corresponds to a lower downside risk).

The outcomes of strategies based on probabilistic forecasts are presented in the two bottom panels, which report the results for methods without and with forecast averaging, respectively. Let us first consider the zero-cost setup. Under these favorable market conditions, only minor differences in trading frequencies can be observed across the considered approaches, confirming the limited impact of the stopping rule on BESS performance. As a result, differences in profitability stem primarily from the selection of charging and discharging hours. Since risk management provides only limited additional gains in this setting, strategies based on probabilistic forecasts do not exhibit clear advantages over their point forecast counterparts. This is reflected in the comparable profit levels, with ARX and ARX-s outperforming the remaining strategies in Spain and ranking among the top three approaches in Germany. Among the probabilistic forecasting methods, MS and MS-ave achieve the highest profit levels.

The picture is different, however, when the setup with the cost of 40 EUR/MWh is considered. In such a case, refraining from trade plays an important role as it allows to limit financial losses. First, it can be observed that strategies based on probabilistic forecasts are more restrictive that ARX-s and reduce the frequencies of trade to around 79.7\%-87.3\% and 54.8\%-63.0\% in Germany and Spain, respectively. At the same time, ARX-s engages in trade on 91.9\% (Germany) and 74.7\% (Spain) of days.

When profit levels are compared, three strategies -- MS-ave, MS, and IS -- stand out. These approaches outperform ARX-s, with MS-ave generating profits that are 2.5\%-5.5\% higher than those achieved by the point forecast. In addition, strategies based on probabilistic forecasts substantially reduce the level of risk. For example, VaR$_{5\%}$ increases from -21.77 and -27.88 to -13.82 and -21.42 in the German and Spanish markets, respectively (values for MS-ave). Hence, in the presence of operational costs, there is a strong evidence that the use of more sophisticated probabilistic strategies yields significant financial benefits.

Additionally, one can observe changes in the optimal level of threshold $q^*$ between the setups with and without the operational costs. In the zero-cost case, the most profitable are the risk-neutral strategies, with $q^*$ oscillating around 0.5. As the operational cost rises, more risk-averse approaches yield higher profits. For example, for the MS-ave strategy, the optimal $q^*$ drops from 0.4 to 0.3 in Germany and 0.6 to 0.4 in Spain.

\begin{figure}
    \centering
    \includegraphics[width=\textwidth]{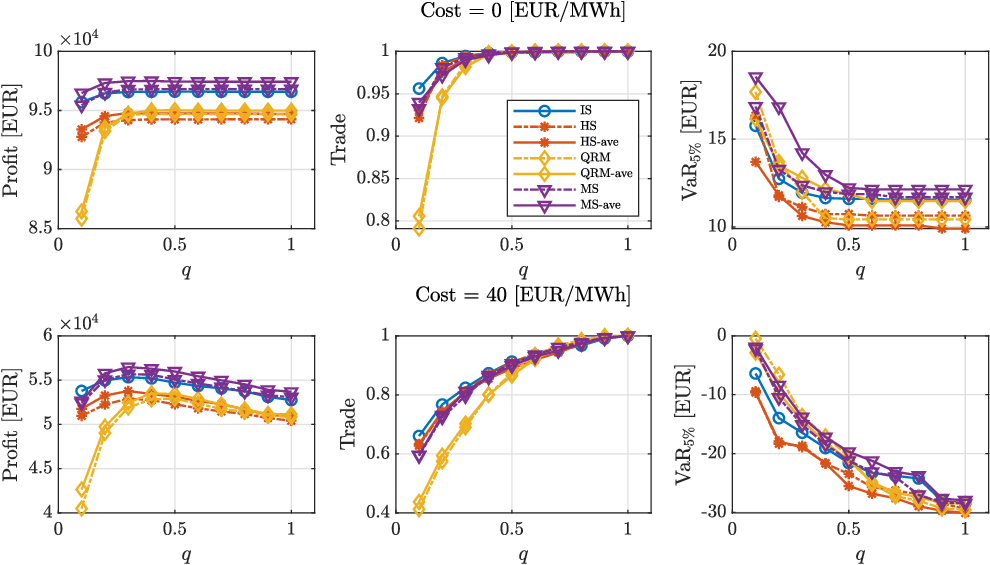}   
    \caption{Financial performance of BESS for German electricity market}
    \label{fig:profit:germ}
\end{figure}

\begin{figure}
    \centering
    \includegraphics[width=\textwidth]{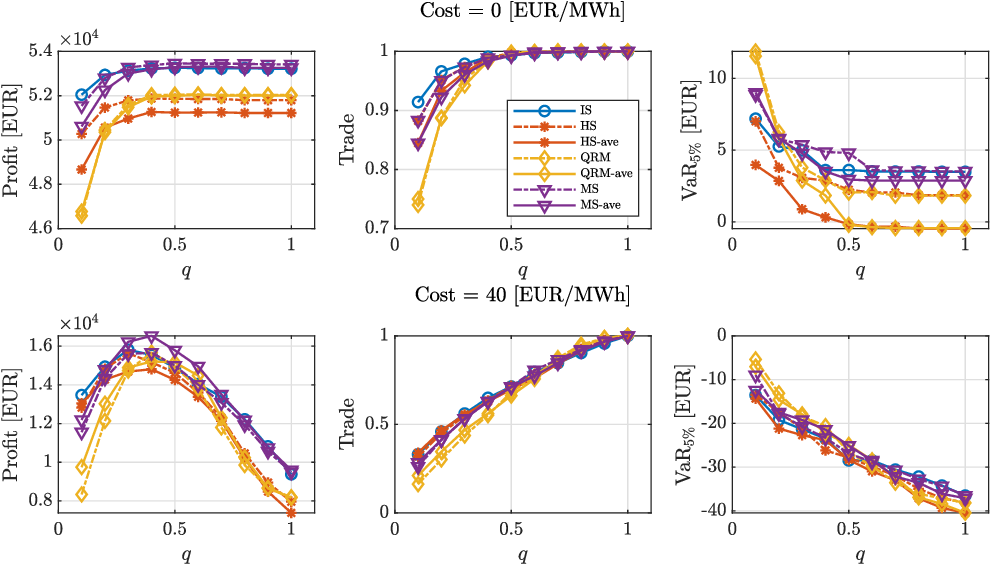}   
    \caption{Financial performance of BESS for Spanish electricity market}
    \label{fig:profit:spain}
\end{figure}

Performance of trading strategies for different levels of risk aversion, represented by $q$, is depicted in Figures \ref{fig:profit:germ}-\ref{fig:profit:spain}. They show profit level (left column), frequency of trade (middle column) and VaR$_{5\%}$ (right column) for setups with $\mathrm{Cost}=0\ \mathrm{EUR/MWh} $ (top row) and $\mathrm{Cost}=40\ \mathrm{EUR/MWh}$ (bottom row). The outcomes confirm that a high level of risk aversion, for instance, $q=0.1$, results in less frequent trade and a lower level of risk. The shape of the profit curves has a clear concave pattern, which becomes more pronounced for higher level of costs. For $\mathrm{Cost}=40\ \mathrm{EUR/MWh}$, the level of profits initially increases and reaches a local maxima for $q$ between 0.3-0.4.

The ranking for forecasting methods is consistent across different risk levels. In Germany, strategies based on MS-ave and MS (violet lines) generate the highest profits for all values of $q$ except 0.1. In Spain, MS-ave or MS outperform the remaining approaches for $q > 0.3$. Consistent with the results reported in Tables~\ref{tab:trade_ger}-\ref{tab:trade_spa}, the third-best performing approach is based on the IS forecasting method. The relative performance of HS- and QRM-based strategies depends on both the market and the level of risk aversion, with the QRM and QRM-ave approaches representing the most conservative strategies.

Finally, it can be observed that the ranking of forecasting methods in terms of BESS profitability is broadly consistent with their forecasting accuracy. The approaches yielding the highest profits (MS-ave, MS, and IS) are also characterized by the lowest CRPS values for profit forecasts (Table~\ref{tab:profit}). In contrast, the QRM-based approaches, which rank among the least profitable strategies, fail to accurately predict BESS profits. This result highlights the importance of capturing the dependence structure between electricity prices across hours. Although the QRM provides relatively well-calibrated marginal distributions of hourly prices, it does not explicitly model cross-hour dependencies. Consequently, it produces less accurate forecasts of profit distributions, which ultimately translates into inferior performance of the corresponding trading strategies.

\begin{table}
\caption{Evaluation of trading strategies -  German market}
\label{tab:trade_ger}
\centering
\begin{tblr}{
colspec={|l|cc|cc|cc|cc|},
cell{1}{2,6} = {c=4}{c},
cell{3}{1} = {c=9}{c},
}
\hline
& Cost = 0 EUR/MWh & & & & Cost = 40 EUR/MWh& & & \\
\hline
& $q^*$ & Trade &  Profit & VaR$_{5\%}$ & $q^*$ & Trade &  Profit & VaR$_{5\%}$ \\
\hline \hline
Germany & & & & & & & & \\
\hline \hline
Oracle	&	--	&	99.91\%	&	107280.27	&	22.51	&	--	&	80.38\%	&	66054.60	&	4.72	\\
\hline
Na\"{\i}ve	&	--	&	100.0\%	&	49.92\%	&	-13.63	&	--	&	100.0\%	&	14.71\%	&	-53.63	\\
ARX	&	--	&	100.0\%	&	90.35\%	&	12.10	&	--	&	100.0\%	&	80.37\%	&	-27.90	\\
ARX-s	&	--	&	99.91\%	&	90.36\%	&	12.47	&	--	&	91.88\%	&	83.03\%	&	-21.77	\\
\hline
IS	&	0.5	&	99.91\%	&	90.04\%	&	11.61	&	0.3	&	82.39\%	&	83.76\%	&	-16.47	\\
HS	&	0.7	&	100.0\%	&	87.86\%	&	10.65	&	0.3	&	81.02\%	&	80.09\%	&	-18.68	\\
QRM	&	0.6	&	100.0\%	&	88.27\%	&	10.43	&	0.5	&	87.32\%	&	80.10\%	&	-20.50	\\
MS	&	0.5	&	99.91\%	&	90.24\%	&	11.87	&	0.3	&	79.65\%	&	84.28\%	&	-15.11	\\
\hline
HS-ave	&	0.3	&	99.18\%	&	88.37\%	&	10.63	&	0.3	&	80.93\%	&	81.39\%	&	-18.92	\\
QRM-ave	&	0.5	&	99.82\%	&	88.58\%	&	11.88	&	0.4	&	80.02\%	&	81.02\%	&	-18.09	\\
MS-ave	&	0.4	&	99.54\%	&	\textbf{90.88\%}	&	\textbf{12.98}	&	0.3	&	81.20\%	&	\textbf{85.49\%}	&	\textbf{-13.82}	\\
\hline
\end{tblr}
\end{table}

\begin{table}
\caption{Evaluation of trading strategies -  Spanish market}
\label{tab:trade_spa}
\centering
\begin{tblr}{
colspec={|l|cc|cc|cc|cc|},
cell{1}{2,6} = {c=4}{c},
cell{3}{1} = {c=9}{c},
}
\hline
& Cost = 0  EUR/MWh& & & & Cost = 40  EUR/MWh& & & \\
\hline
& $q^*$ & Trade &  Profit & VaR$_{5\%}$ & $q^*$ & Trade &  Profit & VaR$_{5\%}$ \\
\hline \hline
Spain & & & & & & & & \\
\hline \hline
Oracle &	--	&	99.18\%	&	61476.70	&	12.60	&	--	&	57.48\%	&	24693.73	&	2.40	\\
\hline
Na\"{\i}ve &	--	&	100.0\%	&	10.97\%	&	-51.05	&	--	&	100.0\%	&	-150.21\%	&	-91.05	\\
ARX	&	--	&	100.0\%	&	87.47\%	&	3.59	&	--	&	100.0\%	&	40.24\%	&	-36.41	\\
ARX-s	&	--	&	99.45\%	&	\textbf{87.56\%}	&	\textbf{4.83}	&	--	&	74.73\%	&	61.43\%	&	-27.88	\\
\hline
IS	&	0.5	&	99.36\%	&	86.63\%	&	3.61	&	0.3	&	56.11\%	&	64.05\%	&	-21.34	\\
HS	&	0.4	&	98.81\%	&	84.38\%	&	2.88	&	0.3	&	54.84\%	&	63.30\%	&	-21.23	\\
QRM	&	0.6	&	99.91\%	&	84.68\%	&	2.08	&	0.4	&	55.20\%	&	63.47\%	&	\textbf{-20.62}	\\
MS	&	0.5	&	99.36\%	&	86.97\%	&	4.80	&	0.4	&	62.96\%	&	63.44\%	&	-23.32	\\
\hline
HS-ave	&	0.4	&	98.81\%	&	83.39\%	&	0.32	&	0.4	&	62.86\%	&	59.97\%	&	-23.84	\\
QRM-ave	&	0.6	&	99.91\%	&	84.65\%	&	-0.37	&	0.4	&	56.11\%	&	61.60\%	&	-22.16	\\
MS-ave	&	0.6	&	99.91\%	&	86.73\%	&	2.88	&	0.4	&	62.86\%	&	\textbf{66.97\%	}&	-21.42	\\
\hline
\end{tblr}
\end{table}

\section{Conclusions}
\label{sec:conclusions}

In this article, we employ ensemble methods to construct probabilistic forecasts of day-ahead electricity prices. The study extends the Multiple Split method proposed by \cite{mac:nit:26} by incorporating forecast averaging and compares its performance with well-estab- lished benchmark approaches, such as In-Sample Errors, Historical Simulation and the Quantile Regression Machine.

An important advantage of ensemble forecasting is its natural extension to multidimensional settings. By jointly calibrating forecast errors across all hours within a day, the proposed approach preserves the correlation structure of the residuals. Consequently, it enables a direct construction of a multidimensional distribution of prices, without the need for explicit modeling of the interdependencies between hours. This property is particularly valuable in BESS management, where operational decisions depend on the joint behavior of electricity prices throughout the day.

The empirical study conducted on the German and Spanish electricity markets demonstrates that ensemble approaches provide accurate and well-calibrated probabilistic forecasts. In particular, the MS and MS-ave methods generate prediction intervals with empirical coverage levels close to their nominal levels, with MS-ave simultaneously achieving (almost always) the lowest CRPS and ES values among the considered approaches. The results also confirm that forecast averaging improves predictive performance across all forecasting frameworks, with reductions in CRPS ranging from 3.7\% to 12.4\%.

The forecasts of electricity prices are then used to construct probabilistic forecasts of BESS profits. The results confirm the superior performance of the newly proposed MS-ave approach. This method provides probabilistic forecasts that are both well calibrated and characterized by the lowest CRPS values among all considered forecasting frameworks. When all possible combinations of charging and discharging hours are considered, the outcomes show that the MS-ave method is followed by the MS, with both approaches outperforming the IS, HS, and QRM methods.

Finally, the economic evaluation demonstrates that accurate probabilistic forecasts can lead to significant financial benefits. The performance of strategies based on the IS, HS, QRM and MS frameworks is compared with ones based on point forecasts (ARX and ARX-s). The results show that under more challenging conditions -- associated with non-zero operational costs -- strategies based on probabilistic forecasts tend to outperform point forecast benchmarks. In particular, the MS-ave approach generates profits that exceed those of ARX-s by 2.5\%-5.5\%, while simultaneously reducing downside risk. For instance, VaR$_{5\%}$ improves from -21.77 EUR to -13.82 EUR in Germany and from -27.88 EUR to -21.42 EUR in Spain. The results also indicate the importance of risk management. While relatively risk-neutral strategies are optimal in the zero-cost setup, more risk-averse behavior leads to higher profits once operational costs are introduced.

Overall, the findings demonstrate that multidimensional probabilistic forecasts can significantly enhance BESS trading decisions by simultaneously increasing profitability and reducing financial risk. The proposed MS-ave framework combines accurate probabilistic forecasting with computational simplicity and flexibility, making it a promising tool for practical energy trading applications.

Several directions for future research remain open. First, the proposed framework may be extended to intraday and balancing markets, where forecasting uncertainty is even greater. Second, alternative point forecasting models, including machine learning and deep learning approaches, could be incorporated into the ensemble construction framework. Finally, future studies could consider more advanced BESS operating constraints, such as battery degradation, multi-cycle operation and participation in multiple electricity markets simultaneously.

\section*{Declaration of generative AI and AI-assisted technologies in the manuscript preparation process}

During the preparation of this work, the authors used ChatGPT and DeepL in order to improve the quality and style of the text. After using these tools, the authors reviewed and edited the content as needed and take full responsibility for the content of the published article.

\section*{CRediT authorship contribution statement}

\textbf{Tomasz Weron:} Conceptualization, Investigation, Software, Visualization, Writing - review \& editing.
\textbf{Katarzyna Maciejowska:} Conceptualization, Funding acquisition, Investigation, Methodology, Writing - original draft, Writing - review \& editing.

\section*{Declaration of competing interest}

The authors declare that they have no known competing financial interests or personal relationships that could have appeared to influence the work reported in this paper.

\section*{Acknowledgments}

This work was partially supported by the National Science Center (NCN), Poland through grant SONATA BIS no. 2019/34/E/HS4/00060.



\begin{thebibliography}{28}
\expandafter\ifx\csname natexlab\endcsname\relax\def\natexlab#1{#1}\fi
\providecommand{\url}[1]{\texttt{#1}}
\providecommand{\path}[1]{#1}
\providecommand{\DOIprefix}{doi:}
\providecommand{\ArXivprefix}{arXiv:}
\providecommand{\URLprefix}{URL: }
\providecommand{\Pubmedprefix}{pmid:}
\providecommand{\doi}[1]{\href{http://dx.doi.org/#1}{\path{#1}}}
\providecommand{\Pubmed}[1]{\href{pmid:#1}{\path{#1}}}
\providecommand{\bibinfo}[2]{#2}
\ifx\xfnm\relax \def\xfnm[#1]{\unskip,\space#1}\fi
\bibitem[{Barber et~al.(2021)Barber, Cand{\`e}s, Ramdas \&
  Tibshirani}]{bar:etal:21}
\bibinfo{author}{Barber, R.~F.}, \bibinfo{author}{Cand{\`e}s, E.~J.},
  \bibinfo{author}{Ramdas, A.}, \& \bibinfo{author}{Tibshirani, R.~J.}
  (\bibinfo{year}{2021}).
\newblock \bibinfo{title}{{Predictive inference with the jackknife+}}.
\newblock {\it \bibinfo{journal}{The Annals of Statistics}\/},  {\it
  \bibinfo{volume}{49}\/}, \bibinfo{pages}{486 -- 507}.
\bibitem[{Billé et~al.(2023)Billé, Gianfreda, {Del Grosso} \&
  Ravazzolo}]{bille:etal:2023}
\bibinfo{author}{Billé, A.~G.}, \bibinfo{author}{Gianfreda, A.},
  \bibinfo{author}{{Del Grosso}, F.}, \& \bibinfo{author}{Ravazzolo, F.}
  (\bibinfo{year}{2023}).
\newblock \bibinfo{title}{Forecasting electricity prices with expert, linear,
  and nonlinear models}.
\newblock {\it \bibinfo{journal}{International Journal of Forecasting}\/},
  {\it \bibinfo{volume}{39}\/}, \bibinfo{pages}{570--586}.
\bibitem[{EIA(2023)}]{eia:23}
\bibinfo{author}{EIA} (\bibinfo{year}{2023}).
\newblock \bibinfo{title}{International {E}nergy {O}utlook}.
\newblock \URLprefix \url{www.eia.gov/aeo}.
\bibitem[{Gneiting et~al.(2007)Gneiting, Balabdaoui \&
  Raftery}]{gne:bal:raf:07}
\bibinfo{author}{Gneiting, T.}, \bibinfo{author}{Balabdaoui, F.}, \&
  \bibinfo{author}{Raftery, A.} (\bibinfo{year}{2007}).
\newblock \bibinfo{title}{Probabilistic forecasts, calibration and sharpness}.
\newblock {\it \bibinfo{journal}{Journal of the Royal Statistical Society
  B}\/},  {\it \bibinfo{volume}{69}\/}, \bibinfo{pages}{243--268}.
\bibitem[{Gneiting et~al.(2008)Gneiting, Stanberry, Grimit, Held \&
  Johnson}]{gne:etal:08}
\bibinfo{author}{Gneiting, T.}, \bibinfo{author}{Stanberry, L.~I.},
  \bibinfo{author}{Grimit, E.~P.}, \bibinfo{author}{Held, L.}, \&
  \bibinfo{author}{Johnson, N.~A.} (\bibinfo{year}{2008}).
\newblock \bibinfo{title}{Assessing probabilistic forecasts of multivariate
  quantities, with an application to ensemble predictions of surface winds}.
\newblock {\it \bibinfo{journal}{TEST}\/},  {\it \bibinfo{volume}{17}\/},
  \bibinfo{pages}{211--235}.
\bibitem[{Hirsch \& Ziel(2026)}]{hir:zie:26}
\bibinfo{author}{Hirsch, S.}, \& \bibinfo{author}{Ziel, F.}
  (\bibinfo{year}{2026}).
\newblock \bibinfo{title}{Probabilistic forecasting for day-ahead electricity
  prices, battery trading strategies and the economic evaluation of predictive
  accuracy}.
\newblock \URLprefix \url{https://arxiv.org/abs/2604.19580}.
\bibitem[{Hubicka et~al.(2019)Hubicka, Marcjasz \& Weron}]{hub:mar:wer:19}
\bibinfo{author}{Hubicka, K.}, \bibinfo{author}{Marcjasz, G.}, \&
  \bibinfo{author}{Weron, R.} (\bibinfo{year}{2019}).
\newblock \bibinfo{title}{A note on averaging day-ahead electricity price
  forecasts across calibration windows}.
\newblock {\it \bibinfo{journal}{IEEE Transactions on Sustainable Energy}\/},
  {\it \bibinfo{volume}{10}\/}, \bibinfo{pages}{321--323}.
\bibitem[{IEA(2026)}]{iea:26}
\bibinfo{author}{IEA} (\bibinfo{year}{2026}).
\newblock \bibinfo{title}{World {E}nergy {O}utlook 2026}.
\newblock \URLprefix
  \url{www.iea.org/reports/global-energy-review-2026}.
\bibitem[{Janczura \& W{\'o}jcik(2022)}]{jan:woj:22}
\bibinfo{author}{Janczura, J.}, \& \bibinfo{author}{W{\'o}jcik, E.}
  (\bibinfo{year}{2022}).
\newblock \bibinfo{title}{Dynamic short-term risk management strategies for the
  choice of electricity market based on probabilistic forecasts of profit and
  risk measures. the german and the polish market case study}.
\newblock {\it \bibinfo{journal}{Energy Economics}\/},  {\it
  \bibinfo{volume}{110}\/}, \bibinfo{pages}{106015}.
\bibitem[{Kath \& Ziel(2021)}]{kath_conformal_2021}
\bibinfo{author}{Kath, C.}, \& \bibinfo{author}{Ziel, F.}
  (\bibinfo{year}{2021}).
\newblock \bibinfo{title}{Conformal prediction interval estimation and
  applications to day-ahead and intraday power markets}.
\newblock {\it \bibinfo{journal}{International Journal of Forecasting}\/},
  {\it \bibinfo{volume}{37}\/}, \bibinfo{pages}{777--799}.
\bibitem[{Koenker \& Hallock(2001)}]{koe:hal:01}
\bibinfo{author}{Koenker, R.}, \& \bibinfo{author}{Hallock, K.~F.}
  (\bibinfo{year}{2001}).
\newblock \bibinfo{title}{Quantile regression}.
\newblock {\it \bibinfo{journal}{Journal of Economic Perspectives}\/},  {\it
  \bibinfo{volume}{15}\/}, \bibinfo{pages}{143--156}.
\bibitem[{Kumbartzky et~al.(2017)Kumbartzky, Schacht, Schulz \&
  Werners}]{kumbartzky_optimal_2017}
\bibinfo{author}{Kumbartzky, N.}, \bibinfo{author}{Schacht, M.},
  \bibinfo{author}{Schulz, K.}, \& \bibinfo{author}{Werners, B.}
  (\bibinfo{year}{2017}).
\newblock \bibinfo{title}{Optimal operation of a {{CHP}} plant participating in
  the {{German}} electricity balancing and day-ahead spot market}.
\newblock {\it \bibinfo{journal}{European Journal of Operational Research}\/},
  {\it \bibinfo{volume}{261}\/}, \bibinfo{pages}{390--404}.
\bibitem[{Kupiec(1995)}]{kup:95}
\bibinfo{author}{Kupiec, P.~H.} (\bibinfo{year}{1995}).
\newblock \bibinfo{title}{Techniques for verifying the accuracy of risk
  measurement models}.
\newblock {\it \bibinfo{journal}{The Journal of Derivatives}\/},  {\it
  \bibinfo{volume}{3}\/}, \bibinfo{pages}{73--84}.
\bibitem[{Lago et~al.(2021)Lago, Marcjasz, De~Schutter \&
  Weron}]{lag:mar:des:wer:21}
\bibinfo{author}{Lago, J.}, \bibinfo{author}{Marcjasz, G.},
  \bibinfo{author}{De~Schutter, B.}, \& \bibinfo{author}{Weron, R.}
  (\bibinfo{year}{2021}).
\newblock \bibinfo{title}{Forecasting day-ahead electricity prices: A review of
  state-of-the-art algorithms, best practices and an open-access benchmark}.
\newblock {\it \bibinfo{journal}{Applied Energy}\/},  {\it
  \bibinfo{volume}{293}\/}, \bibinfo{pages}{116983}.
\bibitem[{Lei et~al.(2018)Lei, G’Sell, Rinaldo, Tibshirani \&
  Wasserman}]{lei:etal:18}
\bibinfo{author}{Lei, J.}, \bibinfo{author}{G’Sell, M.},
  \bibinfo{author}{Rinaldo, A.}, \bibinfo{author}{Tibshirani, R.~J.}, \&
  \bibinfo{author}{Wasserman, L.} (\bibinfo{year}{2018}).
\newblock \bibinfo{title}{Distribution-free predictive inference for
  regression}.
\newblock {\it \bibinfo{journal}{Journal of the American Statistical
  Association}\/},  {\it \bibinfo{volume}{113}\/}, \bibinfo{pages}{1094--1111}.
\bibitem[{Liu et~al.(2017)Liu, Nowotarski, Hong \& Weron}]{liu:now:hon:wer:17}
\bibinfo{author}{Liu, B.}, \bibinfo{author}{Nowotarski, J.},
  \bibinfo{author}{Hong, T.}, \& \bibinfo{author}{Weron, R.}
  (\bibinfo{year}{2017}).
\newblock \bibinfo{title}{Probabilistic load forecasting via {Q}uantile
  {R}egression {A}veraging on sister forecasts}.
\newblock {\it \bibinfo{journal}{IEEE Transactions on Smart Grid}\/},  {\it
  \bibinfo{volume}{8}\/}, \bibinfo{pages}{730--737}.
\bibitem[{Maciejowska(2022)}]{mac:22}
\bibinfo{author}{Maciejowska, K.} (\bibinfo{year}{2022}).
\newblock \bibinfo{title}{Portfolio management of a small {RES} utility with a
  structural vector autoregressive model of electricity markets in {G}ermany}.
\newblock {\it \bibinfo{journal}{Operations Research and Decisions}\/},  {\it
  \bibinfo{volume}{32}\/}, \bibinfo{pages}{75--90}.
\bibitem[{Maciejowska \& Nitka(2026)}]{mac:nit:26}
\bibinfo{author}{Maciejowska, K.}, \& \bibinfo{author}{Nitka, W.}
  (\bibinfo{year}{2026}).
\newblock \bibinfo{title}{Arbitrage in short-term electricity markets: Economic
  value of multidimensional ensemble forecasts}.
\newblock {\it \bibinfo{journal}{Operations Research and Decisions}\/},  {\it
  \bibinfo{volume}{36}\/}. 
\bibitem[{Maciejowska et~al.(2024)Maciejowska, Serafin \&
  Uniejewski}]{mac:uni:ser:24}
\bibinfo{author}{Maciejowska, K.}, \bibinfo{author}{Serafin, T.}, \&
  \bibinfo{author}{Uniejewski, B.} (\bibinfo{year}{2024}).
\newblock \bibinfo{title}{Probabilistic forecasting with a hybrid {Factor-QRA
  approach}: {A}pplication to electricity trading}.
\newblock {\it \bibinfo{journal}{Electric Power Systems Research}\/},  {\it
  \bibinfo{volume}{234}\/}, \bibinfo{pages}{1--12}.
\bibitem[{Marcjasz et~al.(2023)Marcjasz, Narajewski, Weron \&
  Ziel}]{mar:nar:wer:zie:23}
\bibinfo{author}{Marcjasz, G.}, \bibinfo{author}{Narajewski, M.},
  \bibinfo{author}{Weron, R.}, \& \bibinfo{author}{Ziel, F.}
  (\bibinfo{year}{2023}).
\newblock \bibinfo{title}{Distributional neural networks for electricity price
  forecasting}.
\newblock {\it \bibinfo{journal}{Energy Economics}\/},  {\it
  \bibinfo{volume}{125}\/}, \bibinfo{pages}{106843}.
\bibitem[{Marcjasz et~al.(2018)Marcjasz, Serafin \& Weron}]{mar:ser:wer:18}
\bibinfo{author}{Marcjasz, G.}, \bibinfo{author}{Serafin, T.}, \&
  \bibinfo{author}{Weron, R.} (\bibinfo{year}{2018}).
\newblock \bibinfo{title}{Selection of calibration windows for day-ahead
  electricity price forecasting}.
\newblock {\it \bibinfo{journal}{Energies}\/},  {\it \bibinfo{volume}{11}\/},
  \bibinfo{pages}{2364}.
\bibitem[{Marcjasz et~al.(2020)Marcjasz, Uniejewski \& Weron}]{mar:uni:wer:20}
\bibinfo{author}{Marcjasz, G.}, \bibinfo{author}{Uniejewski, B.}, \&
  \bibinfo{author}{Weron, R.} (\bibinfo{year}{2020}).
\newblock \bibinfo{title}{Probabilistic electricity price forecasting with
  {NARX} networks: Combine point or probabilistic forecasts?}
\newblock {\it \bibinfo{journal}{International Journal of Forecasting}\/},
  {\it \bibinfo{volume}{36}\/}, \bibinfo{pages}{466--479}.
\bibitem[{Nowotarski \& Weron(2015)}]{now:wer:15}
\bibinfo{author}{Nowotarski, J.}, \& \bibinfo{author}{Weron, R.}
  (\bibinfo{year}{2015}).
\newblock \bibinfo{title}{Computing electricity spot price prediction intervals
  using quantile regression and forecast averaging}.
\newblock {\it \bibinfo{journal}{Computational Statistics}\/},  {\it
  \bibinfo{volume}{30}\/}, \bibinfo{pages}{791--803}.
\bibitem[{Nowotarski \& Weron(2018)}]{now:wer:18}
\bibinfo{author}{Nowotarski, J.}, \& \bibinfo{author}{Weron, R.}
  (\bibinfo{year}{2018}).
\newblock \bibinfo{title}{Recent advances in electricity price forecasting: {A}
  review of probabilistic forecasting}.
\newblock {\it \bibinfo{journal}{Renewable and Sustainable Energy Reviews}\/},
  {\it \bibinfo{volume}{81}\/}, \bibinfo{pages}{1548--1568}.
\bibitem[{Pinson(2013)}]{pin:13}
\bibinfo{author}{Pinson, P.} (\bibinfo{year}{2013}).
\newblock \bibinfo{title}{Wind energy: {F}orecasting challenges for its
  operational management}.
\newblock {\it \bibinfo{journal}{Statistical Science}\/},  {\it
  \bibinfo{volume}{28}\/}, \bibinfo{pages}{564--585}.
\bibitem[{Serafin et~al.(2019)Serafin, Uniejewski \& Weron}]{ser:uni:wer:19}
\bibinfo{author}{Serafin, T.}, \bibinfo{author}{Uniejewski, B.}, \&
  \bibinfo{author}{Weron, R.} (\bibinfo{year}{2019}).
\newblock \bibinfo{title}{Averaging predictive distributions across calibration
  windows for day-ahead electricity price forecasting}.
\newblock {\it \bibinfo{journal}{Energies}\/},  {\it \bibinfo{volume}{12}\/},
  \bibinfo{pages}{256}.
\bibitem[{Uniejewski \& Maciejowska(2022)}]{uni:mac:22}
\bibinfo{author}{Uniejewski, B.}, \& \bibinfo{author}{Maciejowska, K.}
  (\bibinfo{year}{2022}).
\newblock \bibinfo{title}{Lasso principal component averaging: A fully
  automated approach for point forecast pooling}.
\newblock {\it \bibinfo{journal}{International Journal of Forecasting}\/}, .
\bibitem[{Uniejewski \& Weron(2021)}]{uni:wer:21}
\bibinfo{author}{Uniejewski, B.}, \& \bibinfo{author}{Weron, R.}
  (\bibinfo{year}{2021}).
\newblock \bibinfo{title}{Regularized quantile regression averaging for
  probabilistic electricity price forecasting}.
\newblock {\it \bibinfo{journal}{Energy Economics}\/},  {\it
  \bibinfo{volume}{95}\/}, \bibinfo{pages}{105121}.

\end{thebibliography}

\end{document}